\documentclass[doublecol,linenumbers]{epl2}
\usepackage{epsfig}
\usepackage{dcolumn}
\usepackage{amsmath}
\begin{document}
\title{\bf   Fourier transform analysis of irradiated Weiss oscillations.  }

\author{Jes\'us I\~narrea$^{1,2}$ and Gloria Platero$^2$}

\institute {$^1$Escuela Polit\'ecnica
Superior,Universidad Carlos III,Leganes,Madrid,Spain and
$^2$Unidad Asociada al Instituto de Ciencia de Materiales, CSIC,
Cantoblanco,Madrid,28049,Spain.}

\pacs{nn.mm.xx}{First pacs description}
\pacs{nn.mm.xx}{Second pacs description}
\pacs{nn.mm.xx}{Third pacs description}

\date{\today}
\abstract{
We present a theoretical approach to study the effect of
microwave radiation on the magnetoresistance of a one-dimensional superlattice.
 In our proposal the magnetoresistance of a unidirectional  spatial periodic potential
(superlattice), is
modulated by microwave radiation due to an interference effect
between both, space and time-dependent potentials. The final
magnetoresistance will mainly depend   on the spatial period of the superlattice
 and the radiation frequency.
We consider an  approach to study these effects   based on the fast Fourier transform of the obtained magnetorresistance profiles in function of the inverse of the magnetic field.
Based on this theory we propose the design of a novel radiation sensor for the Terahertz band.}

\maketitle
\section{Introduction}
 Nowadays
magnetotransport properties of highly mobile 2DES are a subject of
increasing interest. In particular the study of the effects that
radiation can produce on these nano-devices is attracting
much attention\cite{ina}. Besides, the interplay of two different
periodic modulations in these physical  systems will give rise to interesting
features in electron dynamics and transport. For instance, the study of the
effect of microwave (MW) radiation on the transport properties of a
 one-dimensional superlattice in the presence of a perpendicular
 magnetic field, ($B$) giving rise to
 Weiss oscillations \cite{weiss,ploog,peeters} represents a topic of great interest. Especially if we
consider that MW radiation gives rise also to magnetoresistance
($R_{xx}$) oscillations in 2DES \cite{mani,zudov,ina2}. Weiss
oscillations are  a type of $R_{xx}$ oscillations observed in
high mobility 2DES with a lateral periodic modulation (superlattice)
imposed in one direction. In this scenario electrons are
subjected simultaneously to two different periodic potentials. One
is space-dependent, i.e., the superlattice, and the other is
time-dependent, i.e., MW radiation. Thus,  interesting properties
are expected to turn up due to the joint effect of both potentials.
\begin{figure}
\centering\epsfxsize=3.5in \epsfysize=2.7in
\epsffile{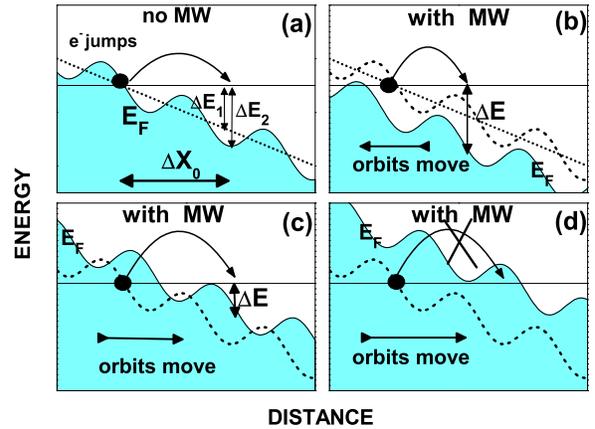} \caption{Schematic diagrams of electronic
transport of a 2DES with a spatial periodic modulation in one
direction.
In Fig.(1.a) no MW field is
present.
When the MW field is on, the orbits
oscillate with $w$. In Fig.(1.b) the orbits move backwards during
the jump. In Fig.(1.c) the orbits are moving forwards but still the
electronic jump can take place. In Fig. (1.d) the orbits are moving
forwards but for a higher MW intensity. Thus the electron movement between orbits
cannot take place because the final state is occupied.
}
\end{figure}

 The origin of
 Weiss oscillations is a
periodically modulated Fermi energy as a direct consequence of the
spatial periodic potential $V(x)$ imposed on the 2DES.
The spatially
modulated Fermi energy, (see Fig. 1), changes dramatically the scattering
conditions between electrons and charged impurities presented in the
sample.
At certain $B$ the
transport will be more dissipative. This corresponds to a peak in
$R_{xx}$\cite{ina3}. Meanwhile in others, the transport will be less
dissipative corresponding to a $R_{xx}$ valley\cite{ina3}. Under this
scenario it is expected that the presence of MW radiation
will alter dramatically the $R_{xx}$ response of the system. We
predict that the combined effect of MW radiation and spatial
potential  will lead the system to an interference regime with
constructive and destructive responses\cite{ina3}. As a consequence $R_{xx}$
will present a modulated profile. The theory presented here can have potential
applications in other fields and systems like nano
electro-mechanical systems (NEMS)\cite{pisto} with AC-potentials and
surfaces acoustic waves (SAW)\cite{saw} in 2DES or quantum dots illuminated
with MW radiation. In other words, the physics presented in this
letter can be of interest for an
audience dealing with the effects that AC or/and DC fields produce
on nano-devices.

In this letter we
consider at the same footing the effects of both  a spatial periodic
modulation and MW radiation on the transport properties of a 2DES
in the presence of $B$. We propose an alternative approach to study the effects of both potentials  based on the fast Fourier transform, (FFT)
\revision{
\cite{anton}},
 of the obtained magnetorresistance profiles in function of the inverse of $B$. We first study the FFT of the system for each potential individually. Then we study jointly the FFT of the system when the two types of potentials are acting simultaneously. We obtain two peaks, corresponding to the two harmonic potentials. An interesting scenario arises when the two FFT peaks coincide. Then we can extract information affecting the two potentials. Also it can be of practical interest, for instance a 2DES with a superlattice of a known spatial period can work as a nanosensor to detect radiation of certain frequencies, for instance for the Terahertz band.

\begin{figure}
\centering\epsfxsize=3.0in \epsfysize=4.5in
\epsffile{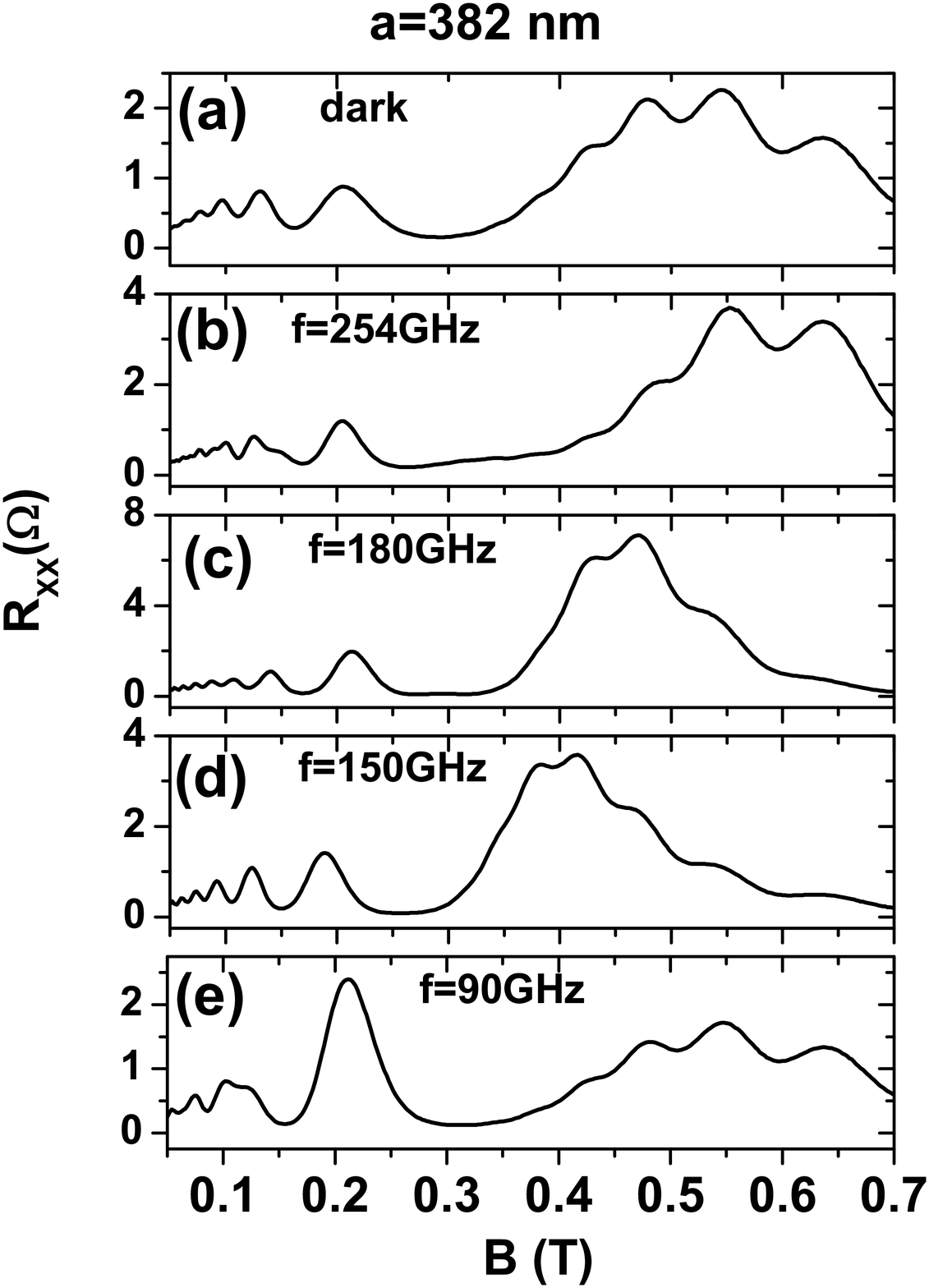} \caption{Calculated $R_{xx}$ as a
function of magnetic field $B$  for dark  and radiation of different
frequencies. The
period of the static modulation is $a=382 nm$ and the modulation
amplitude is $V_{0}\sim 0.1 meV$.T=1K.}
\end{figure}

Our system consists in a 2DES subjected to a perpendicular $B$
(z-direction) and a DC electric field ($E_{dc}$) (x-direction). We
include the unidirectional periodic potential $V(x)=V_{0} \cos Kx$
where $K=2\pi/a$, $a$ being the spatial period of the superlattice.
The total hamiltonian $H$, can be written as:
\begin{eqnarray}
H&=&\frac{P_{x}^{2}}{2m^{*}}+\frac{1}{2}m^{*}w_{c}^{2}(x-X_{0})^{2}-eE_{dc}X_{0}+\frac{1}{2}m^{*}\frac{E_{dc}^{2}}{B^{2}}\nonumber \\
 & &+V_{0} \cos (Kx) =H_{0}+V_{0} \cos (Kx)
\end{eqnarray}
$X_{0}$ is the center of the orbit for the electron spiral motion:
$X_{0}=\frac{\hbar k_{y}}{eB}- \frac{eE_{dc}}{m^{*}w_{c}^{2}}$,
$e$ is the electron charge, $w_{c}$ is the cyclotron frequency.
$H_{0}$ is the hamiltonian of a harmonic quantum oscillator and its
wave functions, the well-known oscillator functions (Hermite
polynomials). We treat $V_{0} \cos (Kx)$ in first order perturbation
theory and the first order energy correction is given by:
$\epsilon_{n}^{(1)}= V_{0} \cos (KX_{0})e^{-X/2}L_{n}(X)=U_{n}\cos
(KX_{0})$, where $X=\frac{1}{2}l^{2}K^{2}$, $L_{n}(X)$ is a Laguerre
polynomial and $l$ the characteristic magnetic length. Therefore the
total energy for the Landau level $n$ is given by:
$\epsilon_{n}=\hbar w_{c}(n+\frac{1}{2})-eE_{dc}X_{0}+U_{n}\cos
(KX_{0})$. This result affects dramatically the Fermi energy as a
function of distance (see Fig. 1), showing now a periodic and tilted
modulation\cite{ina3}.

\begin{figure}
\centering\epsfxsize=3.0in \epsfysize=4.0in
\epsffile{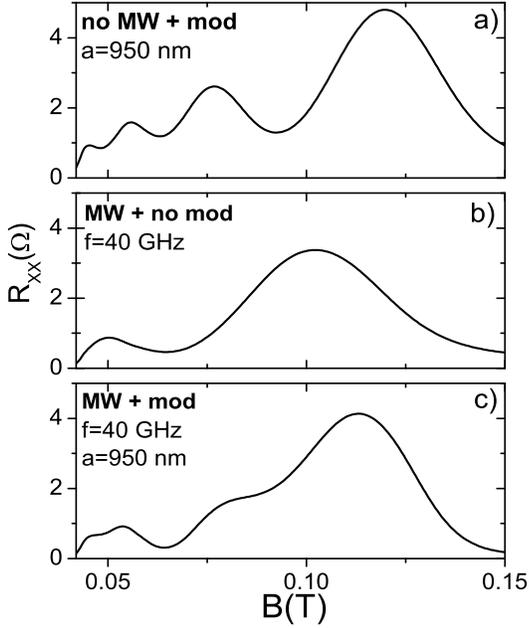} \caption{Calculated $R_{xx}$ vs $B$. In
Fig. (a), we plot the case of superlattice with no radiation. Therefore,
we represent the Weiss oscillations. Fig. (b) represents the case of
a 2DES with no superlattice subjected to radiation. Thus, we
obtain the well-know radiation-induced resistance oscillations.
In Fig. (c) we  represent the joint effect of a one-directional
superlattice with radiation. We obtain a modulated  $R_{xx}$ profile.
The period of the superlattice is 950 nm and the radiation frequency
40 GHz. (T=1K.)}
\end{figure}
\begin{figure}
\centering\epsfxsize=3.5in \epsfysize=5.in
\epsffile{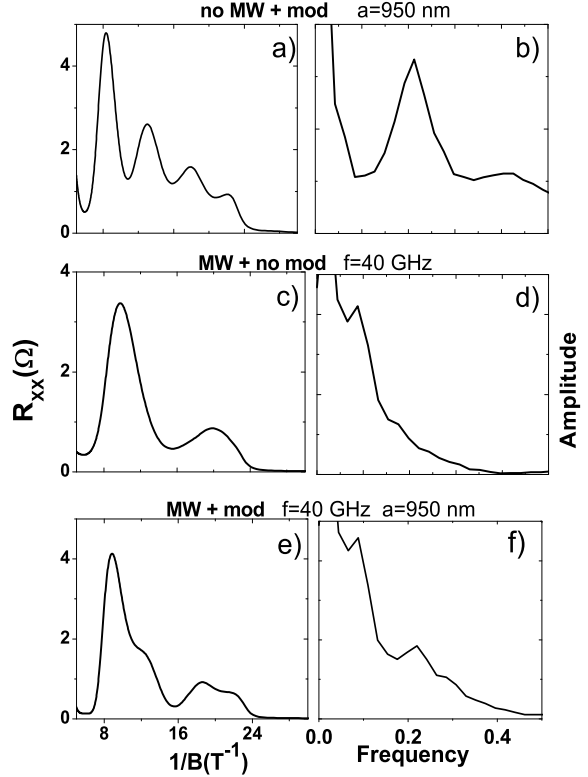} \caption{Left Panels: $R_{xx}$ vs
($1/B)$. In (a) we represent the case of a superlattice of a=950 nm and
no radiation. In (c) we plot the case where the spatial modulation is removed. The frequency of the radiation field is 40 GHz. In (e)
both  spatial and temporal periodic modulations are present.
  In the right panels we represent the FFT of the corresponding left curves. (T=1K).}
\end{figure}
Now we introduce impurity scattering suffered by the electrons in
our model \cite{ina2,ina3,ridley}. When one electron scatters
elastically due to charged impurities, its average orbit center
position changes, in the electric field direction, from $X_{0}$ to
$X_{0}^{'}$. Accordingly, the average advanced distance in the $x$
direction is given by $\Delta X_{0}=X_{0}^{'}-X_{0}\simeq 2R_{c}$,
$R_{c}=\frac{\sqrt{2m^{*}E_{F}}}{eB}$ being the orbit radius.
There is a direct relation between advanced distance and
dissipated energy (see Fig. 1), and when the 2DES is subjected to $V(x)$, the
energy variation in the scattering jump has the expression:
\begin{equation}
\Delta \epsilon\simeq eE_{dc}2R_{c}+U_{n}[\cos (KX_{0})-\cos
(KX_{0}^{'})]
\end{equation}
After some algebra we can obtain from this expression an averaged  energy variation
that can be written as\cite{ina3}:
\begin{equation}
\Delta \epsilon\simeq eE_{dc}2R_{c}- V_{0}e^{-X/2} J_{0}(2\sqrt{X})
\cos\left[ 2\left(R_{c}K+\frac{\pi}{4}\right)\right]
\end{equation}
being $J_{0}$ Bessel function of zero order.
The advanced distance corresponding to $\Delta
\epsilon$ can now be straightforward calculated\cite{}:

\begin{equation}
\Delta X_{T}=2R_{c}- \frac{V_{0}e^{-X/2} J_{0}(2\sqrt{X})}{eE_{dc}}
\cos\left[ 2\left(R_{c}K+\frac{\pi}{4}\right)\right]
\end{equation}

From $\Delta X_{T}$ we calculate a
drift velocity for the electron dissipative transport in the $x$
direction and finally we obtain $R_{xx}$\cite{ina2,ina3,ridley}.
In Fig. 2a we represent
 $R_{xx}$ as a function of $B$. The period of the
static modulation is $a=382$ nm and the modulation amplitude is
$V_{0}\sim 0.1 meV$.
We reproduce Weiss
oscillations with reasonable agreement with
experiments\cite{weiss,ploog}.
\begin{figure}
\centering\epsfxsize=3.0in \epsfysize=4.0in
\epsffile{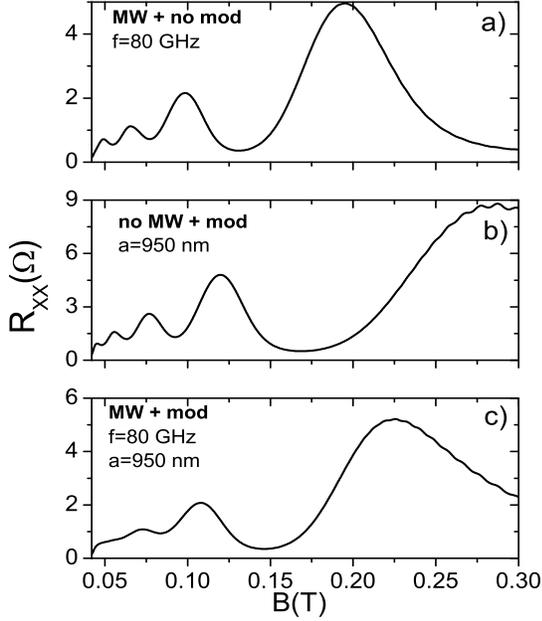} \caption{Calculated $\rho_{xx}$ vs $B$.
Similar panels as in Fig. 3. Now the spatial period of
the superlattice is the same but the radiations frequency is
80 GHz.
 (T=1K.)}
\end{figure}
\begin{figure}
\centering\epsfxsize=3.5in \epsfysize=5.in
\epsffile{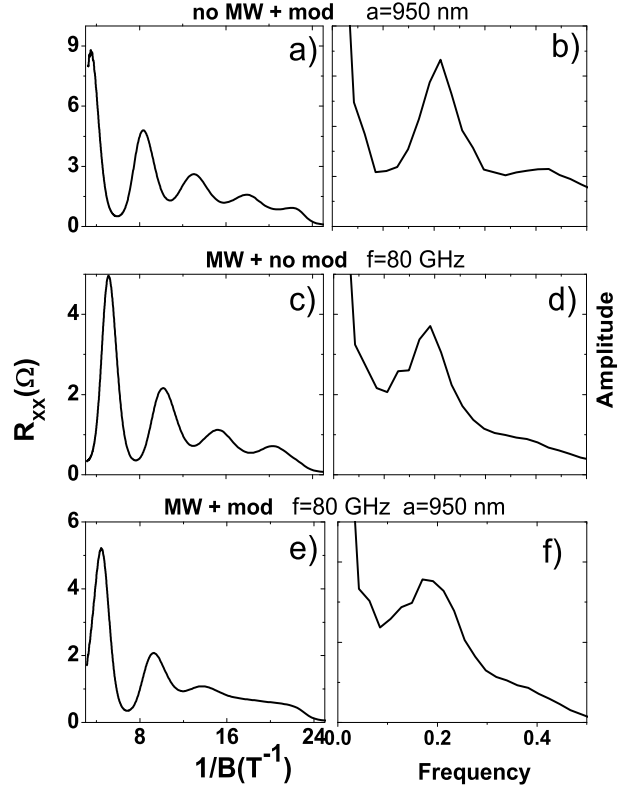} \caption{Calculated $\rho_{xx}$ vs $B$.
Similar panels as in Fig. 4. The spatial period is the same but the
radiation frequency is 80 GHz. (T=1K).}
\end{figure}

If now we switch on the MW field, we expect that its effect will
alter dramatically the $R_{xx}$ response of the system.
\revision{
In our simulations we consider that the MW field is plane polarized
along the $x$-direction that is the direction of the current and the
direction of the spatial potential, i.e., of the superlattice.\cite{anton}} From the
MW driven electron orbits model\cite{ina2,ina3}, we propose that MW radiation forces the
electron orbits to move back and forth at the frequency of MW.
This affects the scattering conditions of the 2DES increasing
or decreasing the distance of the scattering jump giving rise to
MW-induced $R_{xx}$ oscillations. Both effects, periodic spatial
modulation and MW radiation alter simultaneously the average
advanced distance when the electron scatters (see Fig. 1). We obtain
an expression for the total average distance in the $x$ direction\cite{ina3}:
\begin{equation}
\Delta X_{T}=2R_{c}- S\cos\left[
2\left(R_{c}K+\frac{\pi}{4}\right)\right]+A \cos (w \tau)
\end{equation}
where
$S=\frac{V_{0}e^{-X/2} J_{0}(2\sqrt{X})}{eE_{dc}}$
and $A=\frac{e
E_{o}}{m^{*}\sqrt{(w_{c}^{2}-w^{2})^{2}+\gamma^{4}}}$. $E_{o}$ is
the MW electric field amplitude, $\gamma$ is a phenomenologically
introduced\cite{ina2} damping parameter and $\tau$ the scattering
time. From $\Delta X_{T}$ we
calculate an electron drift velocity\cite{ina2,ina3,ridley} and
finally $R_{xx}$ of the system:
\begin{equation}
R_{xx} \propto -S\cos\left[
2\left(R_{c}K+\frac{\pi}{4}\right)\right]+A \cos (w \tau)
\end{equation}

According to this expression we can predict an interference regime
between both periodic potentials which will be reflected on
$R_{xx}$\cite{ina3,ina4} showing a modulated profile when plotted
against $B$.  We can observe this behaviour in Figs. 2b-2c.
In these figures we represent magnetoresistance versus $B$ for
a one-directional superlattice ($a=382 nm$) with MW radiation of different
frequencies: 90, 150, 180 and 254 GHz.
All of them show different modulated profiles due to the interference
effect of both  potentials.
That means that depending mainly on $a$, $w$
and for some values of $B$, $R_{xx}$ will present a constructive
response and a reinforced signal.
For other values of $B$ the
interference will be destructive and the $R_{xx}$ response will
be closer to the one of the dark case.


A further approach to analyze this scenario of irradiated Weiss oscillations is based one the
FFT of the $R_{xx}$ signal versus the inverse of the applied magnetic field. The goal would be
to obtain information from both,  superlattice built on the 2DES  and  radiation. The FFT is carried out on the calculated
data of $R_{xx}$ versus ($1/B$) because they present harmonic distribution; see left panels of Figs. 4 and 6 where
we represent $R_{xx}$ versus the inverse of the magnetic field.
However the standard plots of $R_{xx}$ versus $B$ are not harmonic (see Figs. 3 and 5).
According to equation (6) the spatial part of $R_{xx}$ oscillations depends on the
cosine term: $\cos\left[
2\left(R_{c}K+\frac{\pi}{4}\right)\right]$, where
\begin{equation}
2K R_{c}=\left[2 \frac{2\pi}{a}\frac{\sqrt{2m^{*}E_{F}}}{e}\right] \times \frac{1}{B}=k_{esp}\times \frac{1}{B}
\end{equation}
Here the {\it spatial frequency}, $k_{esp}$,
  is related with the frequency obtained from the  FFT plot, $\nu_{esp}$
 through
\begin{equation}
k_{esp}= 2\pi \nu_{esp}
\end{equation}
(see right panels of Fig. 4).
From this expression we would be able to obtain information
about the spatial period of the superlattice $a$ or even the Fermi level
(electron density) if we previously know the frequency from the FFT graph.

We can apply similar procedure for the temporal part of the $R_{xx}$ oscillations.
According to equation (6) this part depends on $\cos w\tau$, where $\tau=1/W$ is
the scattering time and $W$ the charged impurity scattering rate which in turn can be expressed as\cite{ina5}:
\begin{equation}
W=\left(\frac{e^{4}n_{i}m^{*}}{4\pi\epsilon^{2}\hbar^{3}q_{s}}\right)\left[\frac{1+e^{-\frac{\pi\Gamma}{\hbar w_{c}}}}{1-e^{-\frac{\pi\Gamma}{\hbar w_{c}}}}\right]
\simeq \left(\frac{e^{4}n_{i}m^{*}}{4\pi\epsilon^{2}\hbar^{3}q_{s}}\right)\left[    \frac{2\hbar w_{c}}{\pi\Gamma}\right]
\end{equation}
where $n_{i}$ is the impurity density, $\epsilon$ the dielectric constant, $q_{s}$ the inverse of
the Thomas-Fermi screening length and $\Gamma$ the Landau level width.
Therefore, we can write for $w\tau$, similarly as we did for the spatial part:
\begin{equation}
w\tau = w \times \left(\frac{4\pi\epsilon^{2}\hbar^{3}q_{s}}{  e^{4}n_{i}m^{*}}\right)\left[\frac{\pi\Gamma m^{*}}{2\hbar e}\right]\times\frac{1}{B}
=w_{time}\times\frac{1}{B}
\end{equation}
the  angular frequency $w_{time}$ is related to the frequency of the time dependent FFT $\nu_{time}$
through
\begin{equation}
w_{time}=2\pi\nu_{time}
\end{equation}
(see right panels of Fig. 6).
Thus, knowing $\nu_{time}$ from the FFT plot,  we could obtain information such as the
frequency of radiation or material dependent information, for instance scattering time.

In the above approach we have considered individually the potentials acting on the system.
A more interesting scenario arises when we study jointly the FFT of the system
when acting the two types of potentials simultaneously (Figs. 4f and 6f).
We can have two different situations, in the first one the FFT peaks do not
coincide, they show up far apart (see top and middle panels of Figs. 4 and 6). In this case the obtained information would be as the one
described above. On the other hand, when the two FFT peaks coincide, $ \nu_{time}=\nu_{esp}$ and knowing
previous information of one the potentials, either radiation or superlattice, we could obtain
in turn important information of the other one. An example of
application of the proposed theory it would be the design of radiation sensors; a 2DES with a superlattice of
known spatial period could work as a nanosensor to detect radiation of certain radiation frequency, such as
the Terahertz band. Using the criteria of coincident peaks, we have calculated that to detect, for instance,
radiation of 0.5 THz we would need a superlattice of a spatial period of $a=122$ nm.

In conclusion, we have further examined the problem of irradiated Weiss oscillations.
We have first obtained the interference and modulated profile of $R_{xx}$ vs $B$ arising
from joint effect of superlattice and radiation. Then we have studied this effect from
the perspective of the FFT of $R_{xx}$ response vs $1/B$. We have obtained two peaks
either one corresponding to one potential, time or spatial dependent.
Two scenarios have been obtained, when the FFT peaks positions are the same
and when they do not coincide. The most interesting is the first one, where
 the design of a nanosensor for Terahertz radiation based
on a unidirectional superlattice of known period has been proposed.

This work is supported by
the MINECO (Spain) under grant MAT2011-24331 and ITN
Grant 234970 (EU).

\end{document}